\documentclass[preprint]{aastex61}

\submitjournal{ApJ}

\shorttitle{Solar Microwave Spectrum in the Last Half Century}
\shortauthors{Shimojo et al.}

\begin{document}

\title{Variation of Solar Microwave Spectrum in the Last Half Century}

\correspondingauthor{Masumi Shimojo}
\email{masumi.shimojo@nao.ac.jp}

\author[0000-0002-2350-3749]{Masumi Shimojo}
\affil{National Astronomical Observatory of Japan (NAOJ), National Institutes of Natural Sciences (NINS), Mitaka, Tokyo, 181-8588, Japan}
\affil{Department of Astronomical Science, School of Physical Science, SOKENDAI (The Graduate University of Advanced Studies),  Mitaka, Tokyo, 181-8588, Japan}

\author{Kazumasa Iwai}
\affil{Institute for Space-Earth Environmental Research (ISEE), Nagoya University, Chikusa-ku, Nagoya, 464-8601, Japan}

\author{Ayumi Asai}
\affil{Kwasan and Hida Observatories, Kyoto University, Sakyo-ku, Kyoto, 606-8502, Japan}

\author{Satoshi Nozawa}
\affil{Department of Science, Ibaraki University, Mito, Ibaraki, 310-8512, Japan}

\author{Tetsuhiro Minamidani}
\affil{Nobeyama Radio Observatory, National Astronomical Observatory of Japan (NAOJ), National Institutes of Natural Sciences (NINS), Minamimaki, Minamisaku, Nagano 384-1305, Japan}
\affil{Department of Astronomical Science, School of Physical Science, SOKENDAI (The Graduate University of Advanced Studies),  Mitaka, Tokyo, 181-8588, Japan}

\author{Masao Saito}
\affil{National Astronomical Observatory of Japan (NAOJ), National Institutes of Natural Sciences (NINS), Mitaka, Tokyo, 181-8588, Japan}
\affil{Department of Astronomical Science, School of Physical Science, SOKENDAI (The Graduate University of Advanced Studies),  Mitaka, Tokyo, 181-8588, Japan}

\begin{abstract}
The total solar fluxes at 1, 2, 3.75, and 9.4 GHz were observed continuously from 1957 to 1994 at Toyokawa, and from 1994 until now at Nobeyama, Japan with the current Nobeyama Radio Polarimeters. We examined the multi-frequency and long-term datasets, and found that not only the microwave solar flux but also its monthly standard deviation well indicates the long-term variation of solar activity. Furthermore, we found that the microwave spectra at the solar minima of Cycle 20$\sim$24 agree with each other. These results show that the average atmospheric structure above the upper chromosphere in the quiet Sun has not varied for half a century, and suggest that the energy input for atmospheric heating from the sub-photosphere to the corona has not changed in the quiet Sun despite significantly differing strengths of magnetic activity in the last five solar cycles.

\end{abstract}

\section{Introduction} \label{sec:intro}

The total solar flux\footnote{We have observed the total solar 'flux density' practically. It is commonly, and perhaps inaccurately, referred to as 'flux' in the literature. In the paper, we will use the common term 'flux' rather than 'flux density'.} in microwaves, especially the F10.7 index (total solar flux at 2.8 GHz), is widely used as an indicator of solar activity in the fields of heliophysics and geophysics, because the F10.7 index indicates the variation of solar UV emission in comparison to sunspot number. The time variation in microwaves is traditionally classified into three components based on the timescale of enhancements. The components are a background component from the quiet Sun, a slowly varying component, and a sporadic (burst) component \citep{1965sra..book.....K}. The background component is considered to originate from optically thick thermal bremsstrahlung emission from the atmosphere above the upper chromosphere, but the emission mechanism of the background component is not so simple at lower frequencies (1$\sim$10 GHz). Actually, a simple calculation of thermal bremsstrahlung emission from optically thick layers cannot explain the quiet Sun spectrum measured by \cite{1991ApJ...370..779Z}. The reason for this discrepancy is that the transition region, which is an important contributor in this frequency range, is optically thin \citep[e.g.][]{2011SoPh..273..309S}. The contribution of coronal emission is also a reason for the discrepancy \citep{1991ApJ...370..779Z, 1996ApJ...473..539B}. To reproduce the spectrum of the background component at low frequencies, we need to consider the vertical thermal structure of the quiet Sun from the chromosphere to the corona. Therefore, modeling of the background component is needed, such as those based on recent Radiative-MHD simulations \citep{2004A&A...419..747L}.

The emission sources of a slowly varying component were investigated vigorously using the Westerbork Synthesis Radio Telescope and Very Large Array \citep{1959AnAp...22....1K, 1977A&A....61...79C, 1980A&A....82...30A, 1982ApJ...258..384L, 1982SoPh...80...71C, 1983ApJ...269..698M, 1983A&A...124..103D, 1984ApJ...283..413S, 1987ApJ...315..716W, 1991ApJ...379..366G, 1992ApJS...78..599W, 1994ApJ...426..434A}. From these studies, it is widely accepted that the slowly varying component originates in the thermal bremsstrahlung emission of coronal loops above active regions and thermal gyro-resonance emission from strong magnetic field regions, such as sunspots. A sporadic component is related to flares and coronal mass ejections, and is emitted from the non-thermal electrons accelerated in these phenomena.

The total solar flux at 2.8 GHz has been observed from 1947 \citep{1697669}. Nevertheless, the long-term variation of solar microwave emission itself has not been well investigated except in a few studies \citep[e.g.][]{1969JRASC..63..125C}, because it is hard to study the properties of microwave emission from the single-frequency data. In Japan, the total solar fluxes at 1, 2, 3.75, and 9.4 GHz have been observed continuously from 1957 \citep{1953PRIAN..1..71T, 1957PRIAN..4..60T}, and the dataset is calibrated well with the established method \citep{1973SoPh...29..243T}. \cite{1994SoPh..152..167S, 1995JGR...10019851S} investigated the microwave spectrum of 1980s data obtained in Japan. Because the main purpose of their investigation is the comparison between the solar irradiance measured with the Active Cavity Radiometer Irradiance Monitor (ACRIM) aboard the Solar Maximum Mission (SMM) and microwave spectrum, the period of their investigation is just only 10 years. Hence, there is no studies of  microwave spectrum variation in the period that is longer than one solar cycle yet.
In the paper, we examine the long-term variation of solar microwave emission using the dataset of the last half century, and report the properties of the solar cycle variation in microwaves.

\section{Observation \& Dataset} \label{sec:obs}

In November 1951, monitoring of the total solar flux density at 3.75 GHz started at Toyokawa Observatory, the Research Institute of Atmospherics, Nagoya University \citep{1953PRIAN..1..71T}. Observations at 9.4 GHz started in May 1956, and the 1 and 2 GHz observations started in May and June 1957, respectively at the same site \citep{1957PRIAN..4..60T}. The solar radio telescopes for the 1, 2, and 9.4 GHz observations were prepared for the coordinated observations in the International Geophysical Year lasting from July 1957 to December 1958 \citep{1958BRIA....67T}. For the analysis described in the paper, we need data at all four frequencies, 1, 2, 3.75, and 9.4 GHz. Therefore, we analyzed the data obtained from June 1957.　

In March 1994, the telescopes for the 1, 2, and 9.4 GHz observations were moved to Nobeyama Solar Radio Observatory (NSRO), a branch of the National Astronomical Observatory of Japan (NAOJ), and their operation resumed late in May 1994. At the same time, a new telescope for a 3.75 GHz observation was constructed in Nobeyama. Before the transfer and construction of the telescopes, the 17, 35, and 80 GHz monitoring observations were carried out by NSRO \citep{1985PASJ...37..163N}. Thus, the monitoring observations of the total solar fluxes at 1, 2, 3.75, 9.4, 17, 35, and 80 GHz started at the same site from 1994 May. The monitoring system was named Nobeyama Radio Polarimeters (NoRP). Due to the change of the observing site, the total solar fluxes at 1, 2, and 9.4 GHz were not measured from March to May 1994. Although there is no data loss at 3.75 GHz, we did not analyze the data of this period because no spectrum can be obtained for this study.

NSRO closed on 31 March 2015. Since 1 April 2015, Nobeyama Radio Observatory (NRO), a branch of NAOJ, has continued the operation of NoRP. The dataset used in the paper covers until December 2016.  

NSRO released calibrated daily flux densities at each observing frequency in tabulated form every month at its FTP site\footnote{ftp://solar-pub.nao.ac.jp/pub/nsro/norp/data/daily/}. The calibration method is described in \cite{1973SoPh...29..243T}. In cooperation with the consortium for NoRP scientific operations\footnote{http://solar.nro.nao.ac.jp/norp/html/policy\_new.html}, NRO continues to release the tabulated data using the same FTP site. We used the daily fluxes in the tables for our analysis. The value in the table is not the same as the value calculated using the analysis package of NoRP that is included in the SolarSoftWare \citep[SSW:][]{1998SoPh..182..497F}. The value in the tables is derived after manually removing the effects of radio bursts, bad weather conditions, and instrumental problems. In contrast, when the SSW package is used, the daily flux is automatically calculated from the raw data without such treatment. The difference can be neglected for a study of flares, but the values in the tables are well calibrated and much more reliable for a study of the long-term variation. The total solar flux used in the paper is corrected based on the seasonal variation of the distance between the Sun and the Earth. To compare the solar cycle variation in microwaves with that of the sunspot number, the monthly mean total sunspot number and 13-month smoothed monthly total sunspot number provided from WDC-SILSO, Royal Observatory of Belgium are used \citep{sidc}. 

\section{Solar Cycle Variation in Microwave} \label{sec:flux}

\begin{figure}[h]
\epsscale{.70}
\plotone{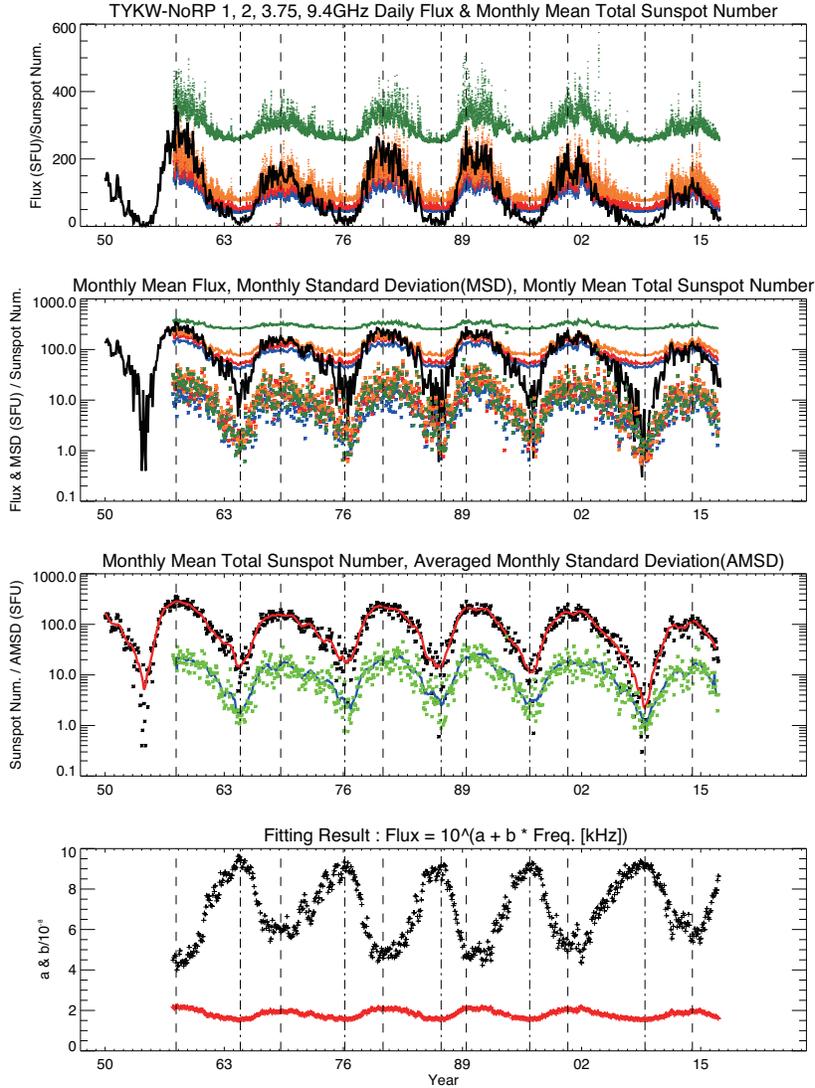}
\caption{The variation of the total solar flux in microwaves. (a) The daily total solar fluxes in microwaves and the monthly mean total sunspot number. (b) The monthly mean microwave fluxes, the monthly mean total sunspot number (solid lines) and the monthly standard deviations (MSD: asterisk). In the two panels, Blue, Red, Orange, Green, and Black indicate 1 GHz, 2 GHz, 3.75 GHz, 9.4 GHz and sunspot number, respectively. (c) The black asterisks indicate the monthly mean total sunspot number, and the red line is the 13-month smoothed monthly total sunspot number. The green asterisks indicate the AMSD, and the blue line is the 13-month smoothed AMSD. (d) The parameters of the model function. The red indicates a: intercept, and the black indicates b: slope. The vertical dashed lines and dotted-dashed lines indicate the solar maxima and solar minima defined from the monthly mean total sunspot number, as shown in Table 1. \label{fig:fig1}}
\end{figure}

Panels (a) and (b) in Figure \ref{fig:fig1} show the long-term variations of the total solar fluxes in microwaves from June 1957 to December 2016. As mentioned in the previous section, the effect of radio bursts is carefully removed. Therefore, the total solar fluxes in panels (a) and (b) in Figure \ref{fig:fig1} are composed of a background component and a slowly varying component, and we can consider that the enhancement around the solar maxima (vertical dashed lines) is caused by the increasing of coronal plasma, and the rising of the number, size, and magnetic field strength of sunspots.

Comparing the daily and monthly total solar fluxes in microwaves and the monthly mean total sunspot number, their variations are very similar at the solar maxima. We can find the counterparts of almost all microwave peaks in the time profile of sunspot number. On the other hand, there are large gaps between them at the solar minima. The total solar fluxes in microwaves do not decrease as significantly as the sunspot number does, because there is a contribution of the emission from the quiet Sun even at solar minima.

To characterize the microwave spectrum in each month, we calculated not only the monthly mean flux but also the monthly standard deviation (MSD) from the daily total solar flux in each frequency (panel (b) in Figure \ref{fig:fig1}). We found that the MSD also shows the solar cycle variation. The MSD around the solar maxima becomes large because the flux from active regions can change up to 40 \% from one day to the next \citep{1978SoPh...56..335F, 1982SoPh...80...71C, 1983A&A...124..103D, 1991ApJ...379..366G}. The emission that causes the large MSD is a slowly varying component, and the variation of the emission strongly depends on the coronal activity and magnetic field strength. At solar minima, no magnetic field is strong enough for gyro-resonance emission, and the coronal activity that enhances the fluctuation of coronal plasma is very low. Consequently, the MSD at the solar minima becomes very small. 

Although the four MSDs can be calculated from the total solar fluxes at the four observing frequencies (1, 2, 3.75, and 9.4 GHz), their variations are quite similar. To simplify our discussion, the averaged value of the four MSDs is used hereafter, and we call the averaged value ``Averaged Monthly Standard Deviation" (AMSD). Panel (c) in Figure \ref{fig:fig1} shows the long-term variation of the AMSD, 13-month smoothed AMSD, monthly mean total sunspot number, and 13-month smoothed monthly total sunspot number. The variation of the AMSD is very similar to that of the monthly mean total sunspot number, and the correlation coefficient between them is 0.71. When we calculate the correlation coefficient between the 13-month smoothed values of the monthly mean total sunspot number and AMSD, the coefficient increases to 0.91. Considering the emission mechanisms at the observing frequencies reviewed at Section \ref{sec:intro}, the AMSD directly reveals the variation of the strong magnetic field regions and the activities in the atmosphere higher than the upper chromosphere. 

Table \ref{tab:tab1} presents the month when the monthly mean total sunspot number or AMSD becomes the maximum/minimum value in each solar cycle. In the table, we can see the difference between the months derived from the monthly mean total sunspot number and that from the AMSD. The difference no doubt results from the differing atmospheric layers that dominate the two indexes, as mentioned above.

\begin{table}[h]
\begin{center}
\begin{tabular}{c c  c c || c c  c c}
\hline \hline
Cycle & Max/Min & SN & AMSD & Cycle & Max/Min & SN & AMSD\\
\hline
19 & Min. &           &                & 22 & Min. & Jun. 1986 & \bf{Sep. 1986} \\
   & Max. & Oct. 1957 & \bf{Jan. 1959} &    & Max. & Jun. 1989 & \bf{Mar. 1989} \\
\hline
20 & Min. & Jul. 1964 & \bf{Sep. 1964} & 23 & Min. & Oct. 1996 & \bf{Jan. 1997} \\
   & Max. & Mar. 1969 & \bf{Jun. 1969} &    & Max. & Jul. 2000 & \bf{Oct. 2003} \\
\hline
21 & Min. & Jul. 1976 & \bf{Jul. 1976} & 24 & Min. & Aug. 2009 & \bf{Jul. 2008} \\
   & Max. & May  1980 & \bf{Jun. 1982} &    & Max. & Feb. 2014 & \bf{Jul. 2014} \\
\hline \hline
\end{tabular}
\end{center}
\caption{The months when the monthly mean total sunspot number (SN) or averaged monthly standard deviation (AMSD) becomes the maximum/minimum value in each solar cycle. \label{tab:tab1}}
\end{table}

\section{Microwave Spectra at Solar Maxima and Minima} \label{sec:spec}

To investigate the solar cycle variation of the microwave spectrum, we carried out the fitting of monthly mean microwave spectra using several simple functions. Finally, we found that an exponential function can give a satisfactory fit to most of the spectra: log$_{10}$(Flux) = a + b $\times ~\nu$, $\nu$ is frequency in Hz. Panel (d) in Figure \ref{fig:fig1} shows the time variation of the fitting model parameters (a: the intercept, b: the slope of the spectrum). The intercept correlates weakly with the solar cycle variation. On the other hand, the slope of the spectrum indicates the anti-correlation. Gyro-resonance emission in lower frequency is stronger than in higher frequency when the emission regions of both frequencies are the same. Because the property of gyro-resonance emission tends to flatten the spectrum at solar maxima, the anti-correlation is a matter of course. We mention that the maximum values of the spectral slopes at the solar minima are about 9.3$\pm$0.3$\times$10$^{-8}$ (3 \% variation) and that the difference is so small though the minimum value of the spectral slope at the solar maximum significantly differs from that in the other solar maxima (5.3$\pm$1.0$\times$10$^{-8}$, $\sim$19 \% variation).

\begin{figure}[h]
\plotone{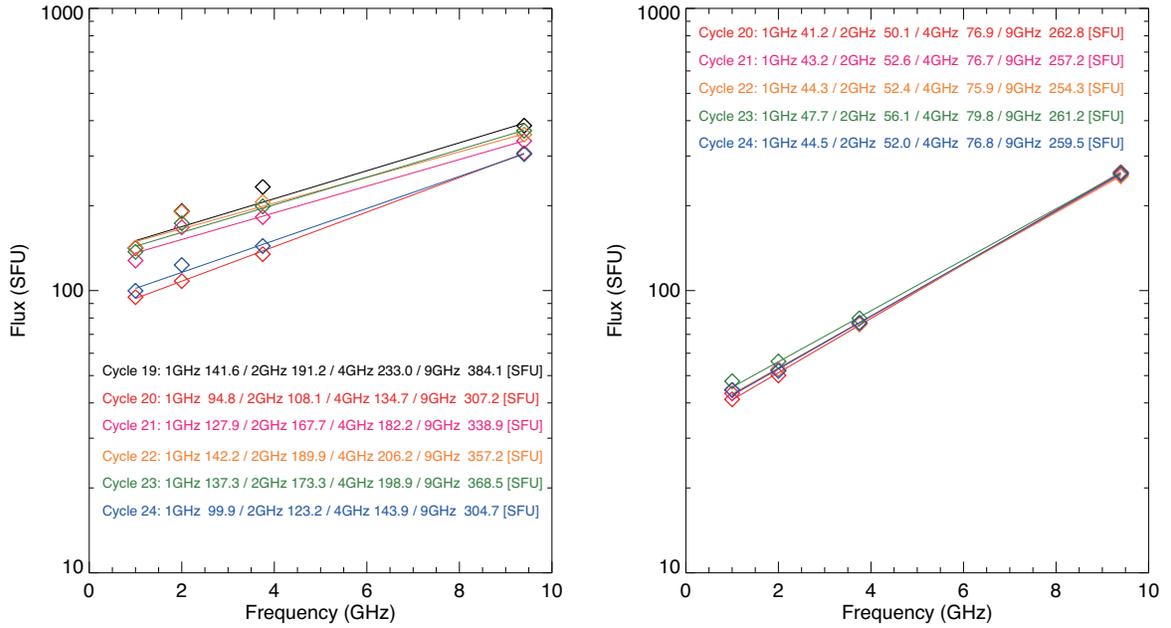}
\caption{ The monthly mean microwave spectra in the months when the AMSD becomes the maximum value (Left panel) or the minimum value (Right panel) in each solar cycle. Color indicates the number of the solar cycle. The months of the solar maxima and minima are shown in bold, Table \ref{tab:tab1}. \label{fig:fig2}}
\end{figure}

To compare the microwave spectra in the solar cycles, we plot the monthly mean solar microwave spectra at the solar maxima and solar minima (Figure \ref{fig:fig2}). To investigate the states where the activity in the upper chromosphere -- corona is highest/lowest, from here, we define ``solar maximum" and ``solar minimum" as the months when the AMSD becomes the maximum/minimum value in each solar cycle. The months are written in bold, Table \ref{tab:tab1}. As shown in the time variation of the fitting model parameter, the spectrum at a solar maximum significantly differs among the cycles. The absolute values of the total solar fluxes at the solar maxima also vary by over 100 \% between the cycles. The difference in the spectra tends to be related to the strength of the solar cycle variation.

A solar minimum is a suitable moment for investigating the spectrum of a background component, because there is no object that emits a slowly varying component and a sporadic component. To obtain the pure spectrum of a background component, we need to select the moment when the coronal activity is extremely low. For this reason, the monthly mean spectrum when the AMSD is lowest in its cycle is suitable for this study. The right panel in Figure \ref{fig:fig2} displays the five monthly mean spectra at the solar minima defined from the AMSD, and the months of the spectra are written in bold, Table \ref{tab:tab1}. The slopes of these spectra are very similar because the variance of the total solar fluxes at the solar minima is only 5.6 SFU (12 \%) at 1 GHz, 4.72 SFU (9 \%) at 2 GHz, 2.23 SFU (3 \%) at 3.75 GHz, and 11.2 SFU (4 \%) at 9 GHz. The values are similar to the precisions of the instruments.

\section{Summary \& Discussion} \label{sec:sum}

We investigated the total solar fluxes at 1, 2, 3.75, and 9.4 GHz obtained from June 1957 to December 2016, and found that the averaged monthly standard deviation (AMSD) of microwave solar fluxes can reveal the long-term variation of solar activity well. Because of the high sensitivity to the activities in the transition region and corona, AMSD can be very useful for investigating the solar-terrestrial environment.

Although an exponential function can give a satisfactory fit to most of the spectra, there is no physical background of the model function. When we execute the fitting of the daily spectra, the chi-square value of the fitting becomes very large around the solar maxima. The tendency can be also seen in the spectra at the solar maxima in Figure \ref{fig:fig2}. The result is consistent with \cite{1994SoPh..152..167S}, and it can be understood as the effect of gyro-resonance emission from sunspots, as mentioned by them, because the spectrum of the emission has a power-law distribution \citep[e.g.][]{1985ARA&A..23..169D}.

To investigate the spectrum of a background component in each cycle, we compared the monthly mean spectra at the solar minima in Cycle 20$\sim$24 defined from the AMSD. Based on the absolute values of the fluxes and the slopes of the spectra, we found that the difference in the spectra is very small. As mentioned in Section \ref{sec:intro}, a background component should reveal the average atmospheric structure of the transition region and lower corona in the quiet Sun. Therefore, the results indicate that the average atmospheric structure above the upper chromosphere in the quiet Sun at solar minima, which may be related to the energy input for atmospheric heating from the sub-photosphere to the corona, has not varied for half a century. The peaks of sunspot numbers at the solar maxima varied during the period. In other words, the Sun’s magnetic activity has been changing significantly \citep[e.g.][]{2012ApJ...757L...8L, 2012SoPh..281..577P, 2012ApJ...750L..42G, 2012ApJ...753..157S, 2013ApJ...763...23S, 2013PASJ...65S..16S}. Nevertheless, our observing results suggest that the energy input for atmospheric heating from the sub-photosphere to the corona has not changed in the quiet Sun. It is commonly supposed that the energy input in the quiet Sun is caused by convection motion and the local dynamo mechanism in the sub-photosphere. Therefore, our results may indicate that the properties of convection motion and the local dynamo mechanism have not changed for half a century, and the solar cycle differences in strong magnetic field generated by the global dynamo mechanism.

\acknowledgments
The authors would like to express thank to the people who have participated in the operation of the solar radio telescopes at Toyokawa and Nobeyama Radio Polarimeters for over half a century.  The Nobeyama Radio Polarimeters (NoRP) are operated by Nobeyama Radio Observatory, a branch of National Astronomical Observatory of Japan, and their observing data are verified scientifically by the consortium for NoRP scientific operations. This work was carried out on the Solar Data Analysis System operated by the Astronomy Data Center of the National Astronomical Observatory of Japan. M.S. was supported by JSPS KAKENHI Grant Number JP17K05397. A.A. was supported by JSPS KAKENHI Grant Numbers JP15H05816 and JP15K17772.

\facility{NoRP}


\begin{thebibliography}{}

\bibitem[{{Alissandrakis} {et~al.}(1980){Alissandrakis}, {Kundu}, \&
  {Lantos}}]{1980A&A....82...30A}
{Alissandrakis}, C.~E., {Kundu}, M.~R., and {Lantos}, P. 1980, \aap, 82, 30

\bibitem[{{Aschwanden} \& {Bastian}(1994)}]{1994ApJ...426..434A}
{Aschwanden}, M.~J., and {Bastian}, T.~S. 1994, \apj, 426, 434

\bibitem[{{Bastian} {et~al.}(1996){Bastian}, {Dulk}, \&
  {Leblanc}}]{1996ApJ...473..539B}
{Bastian}, T.~S., {Dulk}, G.~A., and {Leblanc}, Y. 1996, \apj, 473, 539

\bibitem[{{Chiuderi-Drago} {et~al.}(1982){Chiuderi-Drago}, {Bandiera},
  {Willson}, {Slottje}, {Falciani}, {Antonucci}, {Lang}, and
  {Shibasaki}}]{1982SoPh...80...71C}
{Chiuderi-Drago}, F., {Bandiera}, R., {Willson}, R.~F., {et~al.} 1982,
  \solphys, 80, 71

\bibitem[{{Chiuderi-Drago} {et~al.}(1977){Chiuderi-Drago}, {Felli}, and
  {Tofani}}]{1977A&A....61...79C}
{Chiuderi Drago}, F., {Felli}, M., \& {Tofani}, G. 1977, \aap, 61, 79

\bibitem[{Covington(1948)}]{1697669}
Covington, A. 1948, Proceedings of the IRE, 36, 454

\bibitem[{{Covington}(1969)}]{1969JRASC..63..125C}
{Covington}, A.~E. 1969, \jrasc, 63, 125


\bibitem[{{Dulk} \& {Gary}(1983)}]{1983A&A...124..103D}
{Dulk}, G.~A., and {Gary}, D.~E. 1983, \aap, 124, 103

\bibitem[{{Dulk}(1985)}]{1985ARA&A..23..169D}
{Dulk}, G.~A. 1985, \araa, 23, 169

\bibitem[{{Falchi} {et~al.}(1978){Falchi}, {Felli}, {Pampaloni}, and
  {Tofani}}]{1978SoPh...56..335F}
{Falchi}, A.~D., {Felli}, M., {Pampaloni}, P., and {Tofani}, G. 1978, \solphys,
  56, 335

\bibitem[{{Freeland} \& {Handy}(1998)}]{1998SoPh..182..497F}
{Freeland}, S.~L., and {Handy}, B.~N. 1998, \solphys, 182, 497

\bibitem[{{Gopalswamy} {et~al.}(1991){Gopalswamy}, {White}, \&
  {Kundu}}]{1991ApJ...379..366G}
{Gopalswamy}, N., {White}, S.~M., and {Kundu}, M.~R. 1991, \apj, 379, 366

\bibitem[{{Gopalswamy} {et~al.}(2012){Gopalswamy}, {Yashiro}, {M{\"a}kel{\"a}},
  {Michalek}, {Shibasaki}, and {Hathaway}}]{2012ApJ...750L..42G}
{Gopalswamy}, N., {Yashiro}, S., {M{\"a}kel{\"a}}, P., {et~al.} 2012, \apjl,
  750, L42

\bibitem[{{Kundu}(1959)}]{1959AnAp...22....1K}
{Kundu}, M.~R. 1959, Annales d'Astrophysique, 22, 1

\bibitem[{{Kundu}(1965)}]{1965sra..book.....K}
{Kundu}, M.~R. 1965, {Solar radio astronomy} (New York: Interscience Publication, 1965)

\bibitem[{{Lang} {et~al.}(1982){Lang}, {Willson}, \&
  {Rayrole}}]{1982ApJ...258..384L}
{Lang}, K.~R., {Willson}, R.~F., and {Rayrole}, J. 1982, \apj, 258, 384

\bibitem[{{Livingston} {et~al.}(2012){Livingston}, {Penn}, \&
  {Svalgaard}}]{2012ApJ...757L...8L}
{Livingston}, W., {Penn}, M.~J., and {Svalgaard}, L. 2012, \apjl, 757, L8

\bibitem[{{Loukitcheva} {et~al.}(2004){Loukitcheva}, {Solanki}, {Carlsson}, and
  {Stein}}]{2004A&A...419..747L}
{Loukitcheva}, M., {Solanki}, S.~K., {Carlsson}, M., and {Stein}, R.~F. 2004,
  \aap, 419, 747

\bibitem[{{McConnell} \& {Kundu}(1983)}]{1983ApJ...269..698M}
{McConnell}, D., and {Kundu}, M.~R. 1983, \apj, 269, 698

\bibitem[{{Nakajima} {et~al.}(1985){Nakajima}, {Sekiguchi}, {Sawa}, {Kai}, and
  {Kawashima}}]{1985PASJ...37..163N}
{Nakajima}, H., {Sekiguchi}, H., {Sawa}, M., {Kai}, K., and {Kawashima}, S.
  1985, \pasj, 37, 163

\bibitem[{{Petrie}(2012)}]{2012SoPh..281..577P}
{Petrie}, G.~J.~D. 2012, \solphys, 281, 577

\bibitem[{{Schmahl} \& {Kundu}(1994)}]{1994SoPh..152..167S} 
{Schmahl}, E.~J., \& {Kundu}, M.~R.\ 1994, \solphys, 152, 167 

\bibitem[{{Schmahl} \& {Kundu}(1995)}]{1995JGR...10019851S} 
{Schmahl}, E.~J., \& {Kundu}, M.~R.\ 1995, \jgr, 100, 19851 

\bibitem[{{Shevgaonkar} \& {Kundu}(1984)}]{1984ApJ...283..413S}
{Shevgaonkar}, R.~K., \& {Kundu}, M.~R. 1984, \apj, 283, 413

\bibitem[{{Shibasaki} {et~al.}(2011){Shibasaki}, {Alissandrakis}, and
  {Pohjolainen}}]{2011SoPh..273..309S}
{Shibasaki}, K., {Alissandrakis}, C.~E., and {Pohjolainen}, S. 2011, \solphys,
  273, 309

\bibitem[{{Shimojo}(2013)}]{2013PASJ...65S..16S}
{Shimojo}, M. 2013, \pasj, 65, 16

\bibitem[{{Shiota} {et~al.}(2012){Shiota}, {Tsuneta}, {Shimojo}, {Sako},
  {Orozco Su{\'a}rez}, \& {Ishikawa}}]{2012ApJ...753..157S}
{Shiota}, D., {Tsuneta}, S., {Shimojo}, M., {et~al.} 2012, \apj, 753, 157

\bibitem[{{SILSO World Data Center}(1949-2016)}]{sidc}
{SILSO World Data Center}. 1949-2016, International Sunspot Number Monthly
  Bulletin and online catalogue, http://sidc.oma.be/silso/

\bibitem[{{Svalgaard} \& {Kamide}(2013)}]{2013ApJ...763...23S}
{Svalgaard}, L., and {Kamide}, Y. 2013, \apj, 763, 23

\bibitem[{{Tanaka} {et~al.}(1953){Tanaka}, {Kakinuma}, {Jindoh}, \&
  {Takayanagi}}]{1953PRIAN..1..71T}
{Tanaka}, H., {Kakinuma}, T., {Jindoh}, H., and {Takayanagi}, T. 1953,
  Proceedings of the Research Institute of Atmospherics, Nagoya University, 1,
  71

\bibitem[{{Tanaka} \& {Kakinuma}(1957)}]{1957PRIAN..4..60T}
{Tanaka}, H., and {Kakinuma}, T. 1957, Proceedings of the Research Institute of
  Atmospherics, Nagoya University, 4, 60

\bibitem[{{Tanaka} {et~al.}(1958){Tanaka}, {Kakinuma}, {Jindoh}, {Takayanagi},
  \& {Torii}}]{1958BRIA....67T}
{Tanaka}, H., {Kakinuma}, T., {Jindoh}, H., {Takayanagi}, T., and {Torii}, C.
  1958, Bulletin of the Research Institute of Atmospherics, Nagoya University,
  8, 67

\bibitem[{{Tanaka} {et~al.}(1973){Tanaka}, {Castelli}, {Covington},
  {Kr{\"u}ger}, {Landecker}, \& {Tlamicha}}]{1973SoPh...29..243T}
{Tanaka}, H., {Castelli}, J.~P., {Covington}, A.~E., {et~al.} 1973, \solphys,
  29, 243
  
\bibitem[{{Webb} {et~al.}(1987){Webb}, {Holman}, {Davis}, {Kundu}, and
  {Shevgaonkar}}]{1987ApJ...315..716W}
{Webb}, D.~F., {Holman}, G.~D., {Davis}, J.~M., {Kundu}, M.~R., and
  {Shevgaonkar}, R.~K. 1987, \apj, 315, 716

\bibitem[{{White} {et~al.}(1992){White}, {Kundu}, and
  {Gopalswamy}}]{1992ApJS...78..599W}
{White}, S.~M., {Kundu}, M.~R., and {Gopalswamy}, N. 1992, \apjs, 78, 599

\bibitem[{{Zirin} {et~al.}(1991){Zirin}, {Baumert}, and
  {Hurford}}]{1991ApJ...370..779Z}
{Zirin}, H., {Baumert}, B.~M., and {Hurford}, G.~J. 1991, \apj, 370, 779

\end{thebibliography}
\end{document}